\documentclass[journal,a4paper]{IEEEtran}
\usepackage[utf8]{inputenc}
\usepackage{amsmath}
\usepackage{mathtools}
\usepackage{graphicx}
\usepackage{float}
\usepackage{pgfplots}
\usepackage{tikz}
\usetikzlibrary{plotmarks}
\usetikzlibrary{external}
\usetikzlibrary{arrows}
\usetikzlibrary{shapes,decorations}
\usetikzlibrary{colorbrewer}
\usepackage{caption}
\usepackage{cite}
\pgfplotsset{compat=1.14}
\usepackage{xcolor}
\usetikzlibrary{shapes}
\usetikzlibrary{plotmarks}
\usetikzlibrary{external}
\usetikzlibrary{arrows}
\usetikzlibrary{shapes,decorations}
\usetikzlibrary{colorbrewer}
\usepackage{caption}
\pgfplotsset{compat=1.14}
\usepackage[utf8]{inputenc}
\newcommand{\rom}[1]{\uppercase\expandafter{\romannumeral #1\relax}}
\usepackage{amssymb}

\begin{document}

\title{The Gaussian Noise Model in the Presence of Inter-channel Stimulated Raman Scattering}
%
%

\author{Daniel~Semrau,~\IEEEmembership{Student Member,~IEEE,} and~Polina~Bayvel,~\IEEEmembership{Fellow,~IEEE,~Fellow,~OSA}
\thanks{This work was supported by a UK EPSRC programme grant UNLOC (EP/J017582/1) and a Doctoral Training Partnership (DTP) studentship for Daniel Semrau.}
\thanks{D. Semrau and P. Bayvel are with the Optical Networks Group, University College London, London
WC1E 7JE, U.K. (e-mail: \{uceedfs; p.bayvel\}@ucl.ac.uk.)}
}

\maketitle

\markboth{\today}%
{}

\begin{abstract}
A Gaussian noise (GN) model is presented that properly accounts for an arbitrary frequency dependent signal power profile along the link. This enables the evaluation of the impact of inter-channel stimulated Raman scattering (ISRS) on the optical Kerr nonlinearity. Additionally, the frequency dependent fiber attenuation can be taken into account and transmission systems that use hybrid amplification schemes can be modeled, where distributed Raman amplification is partly applied over the optical spectrum. To include the latter two cases, a set of coupled ordinary differential equations must be numerically solved in order to obtain the signal power profile yielding a semi-analytical model. However for lumped amplification and negligible variation in fiber attenuation, a less complex and fully analytical model is presented which is referred to as the ISRS GN model. The derived model is exact to first-order for Gaussian modulated signals and extensively validated by numerical split-step simulations. A maximum deviation of $0.1$~dB in nonlinear interference power between simulations and the ISRS GN model is found. The model is applied to a transmission system that occupies an optical bandwidth of $10$~THz, representing the entire C+L band. At optimum launch power, changes of up to $2$~dB in nonlinear interference power due to ISRS are reported. Furthermore, comparable models published in the literature are benchmarked against the ISRS GN model.
\end{abstract}


\IEEEpeerreviewmaketitle

\section{Introduction}

\IEEEPARstart{A}{nalytical} models that predict the performance degradation in optical fiber communications due to Kerr nonlinearity have enjoyed significant popularity in recent years. Most approaches analytically solve the nonlinear Schr\"odinger equation using a first-order perturbation approach with respect to the nonlinearity coefficient. The resulting expressions offer unique insight into the underlaying parameter dependencies and are key enablers for efficient system design \cite{Hasegawa_2017_ofd}, rapid achievable rate estimations of point-to-point links \cite{Semrau_2016_air,Shevchenko_2016_air,Bosco_2011_aro} and physical layer aware network optimization. The latter is essential for optical network abstraction and virtualization leading to optimal and intelligent techniques to maximize optical network capacity \cite{Anagnostopoulos_2007_pli}. Analytical models also offer a significant reduction in computational complexity with minor inaccuracies compared to split-step simulations and experiments \cite{Saavedra_2017_eao,Nespola_2014_gvo,Nespola_2015_evo,Galdino_2016_edo,Poggiolini_2015_asa,Semrau_2017_ace}.
\par
The literature offers a wide range of analytical models varying in accuracy and complexity \cite{Splett_1993_utc,Tang_2002_tcc,Poggiolini_2012_tgm,Chen_2010_cef,Johannisson_2013_pao,Mecozzi_2012_nsl,Secondini_2012_afc,Dar_2013_pon,Carena_2014_emo,Golani_2016_mtb,Serena_2015_ate,Ghazisaeidi_2017_ato}. 
The first approaches in the context of modern coherent receivers and dispersion uncompensated links date back to 1993 and 2002 \cite{Splett_1993_utc,Tang_2002_tcc}, enabling the computation of the perturbation caused by Kerr nonlinearity. Similar results were independently derived by other groups and the model became widely known as the Gaussian noise (GN) model \cite{Poggiolini_2012_tgm,Carena_2012_mot,Chen_2010_cef,Johannisson_2013_pao,Serena_2015_ate}. A key assumption of the works is the signal Gaussianity assumption, which is that the signal can be written as a Gaussian process at the fiber input. As a result of this assumption, the GN model relies on large accumulated dispersion \cite[Sec. 6]{Carena_2014_emo} and signals with high cardinality \cite[Sec. 4]{Dar_2013_pon}. Two conditions that are satisfied in most cases of modern coherent fiber communication.
\par 
The popularity of the GN model undoubtedly origins in its moderate complexity. However, as a result it fails to predict certain properties of nonlinear interference such as modulation format dependence \cite{Mecozzi_2012_nsl,Dar_2013_pon,Secondini_2012_afc}, symbol rate dependence \cite{Shieh_2008_coo,Bononi_2013_pdo,Du_2011_pts,Zhuge_2012_coi,Poggiolini_2016_ane}, nonlinear phase noise \cite{Dar_2013_pon,Dar_2014_aon}, long temporal correlations \cite{Secondini_2012_afc,Dar_2013_pon} and the dependence on the memory length of the fiber-optic channel \cite{Agrell_2014_coa}. In order to account for those properties, significantly more complex models have been proposed \cite{Mecozzi_2012_nsl,Secondini_2012_afc,Dar_2013_pon,Carena_2014_emo,Golani_2016_mtb,Agrell_2014_coa}. Comprehensive overviews can be found in \cite{Dar_2015_ini,Poggiolini_2017_rai}. 
\par
The impact of the mentioned properties are usually small for lumped, dispersion unmanaged, multi-span systems that use high-order modulation formats and the GN model can be considered sufficiently accurate. Recently, the conventional GN model was experimentally validated for the central channel and optical bandwidths up to $7.3$~THz with a deviation of only $0.4$~dB in nonlinear interference (NLI) power \cite{Saavedra_2017_eio,Saavedra_2017_eao}. 
\par 
An assumption of all above-mentioned works is that every frequency component experiences the same power evolution along the link. They are therefore inaccurate in the prediction of ultra-wideband transmission systems where the variation of the fiber attenuation is not negligible and for bandwidths where inter-channel stimulated Raman scattering (ISRS) is significant.
\par 
Inter-channel stimulated Raman scattering (ISRS) is a non-parametric nonlinear process that amplifies low frequency components at the expensive of high frequency components within the same optical signal. In modern optical communications that use coherent technology in combination with high dispersive links, ISRS effectively introduces a different power profile for each frequency component \cite{Forghieri_1995_eom,Ho_2000_spo,Norimatsu_2001_wdd}. For C band transmission (approximately $5$~THz), as defined from the availability of the erbium doped fiber amplifier (EDFA), ISRS is not significant and its impact is negligible in most cases. However, for systems that occupy the entire C+L band (approximately $10$~THz) or beyond, ISRS becomes significant and it must be taken into account. Beyond C+L band transmission could be enabled by lumped Raman amplification or the use of other dopants such as bismuth \cite{Dianov_2013_aie} or quantum dots in semiconductor amplifiers \cite{Renaudier_2017_f1c} \cite{Agrell_2016_roo}.
\par
The first approach to include ISRS in the GN model was published in \cite{Semrau_17_ard}. The authors approximated the signal power profile with an exponential decay using an effective attenuation coefficient that matches the (frequency dependent) effective length in the presence of ISRS. Although first conclusions on the impact of ISRS could be drawn, this method exhibits two shortcomings. First, when ISRS is significant, the resulting signal power profile does not resemble an exponential decay, particularly in the beginning of the fiber span, where Kerr nonlinearity prevails. Second, the work assumes that, during the four-wave mixing (FWM) process, every participating frequency component attenuates in the same manner as the channel of interest. In other words, for the induced nonlinear perturbation at frequency $f$, every frequency component in the triplet $\left(f,f_1,f_2\right)$ attenuates as $f$ during the nonlinear mixing. This does not accurately represent the FWM process, as each frequency component attenuates in a different manner.
\par 
A more rigorous approach to include ISRS in the GN model was published in \cite{Roberts_17_cpo,Cantono_2012_ada}. In both works, the channel under test (i.e. frequency component $f$) attenuates precisely according to the signal power profile $g\left(f\right)$ resulting from ISRS (or any arbitrary profile), lifting the exponential decay assumption in \cite{Semrau_17_ard}. However, for an attenuation profile that is linear in frequency (like the one resulting from ISRS), the frequencies in the triplet $\left(f,f_1,f_2\right)$ also attenuate according to $f$ during the FWM process. Therefore, this approach overestimates the impact of ISRS on the Kerr nonlinearity and the frequency dependent signal power profile is not accurately taken into account.
\par \vspace{\baselineskip}
In this paper, a Gaussian noise model is presented that properly accounts for any arbitrary frequency dependent signal power profile. This enables the modeling of nonlinear interference in ultra-wideband regimes where ISRS is significant. The model is referred to as the ISRS GN model. Additionally, the variation in fiber attenuation and hybrid amplification schemes can be included. In general, the signal power profile is obtained by numerically solving a set of coupled ordinary differential equations (ODE). This yields a semi-analytical ISRS GN model which relies on a numerical ODE solver. 
\par 
However, for lumped amplification and negligible variation in fiber attenuation, a fully analytically model is derived based on a linear approximation on the ISRS gain function. This reduces model as well as computational complexity. The analytical ISRS GN model holds for bandwidths up to approximately $15$~THz after which the Raman gain function cannot be considered linear anymore.
\par 
This paper is organized as follows. In Sec. \ref{sec:TheISRSGNmodel}, the ISRS GN model is presented and its key derivation steps are briefly outlined. The detailed derivation can be found in the Appendix \ref{derivation}. The model is extensively validated by split-step simulations in Sec. \ref{sec:validation} and applied to a C+L band transmission system based on standard single mode fiber (SMF) spans in Sec. \ref{sec:C+Lband}. In Sec. \ref{sec:validation} and \ref{sec:C+Lband}, the results in \cite{Semrau_17_ard,Roberts_17_cpo,Cantono_2012_ada} are benchmarked against the ISRS GN model.
\par 
\section{The ISRS GN model}
\label{sec:TheISRSGNmodel}
In order to maximize the information throughput of an optical communication system, it is vital to evaluate and maximize the performance of each individual channel that is transmitted. After coherent detection and electronic dispersion compensation, the channel dependent signal-to-noise ratio (SNR) can be calculated as
\begin{equation}
\begin{split}
\label{eq:SNR}
\textnormal{SNR}_i \approx \frac{P_i}{nP_\textnormal{ASE} + \eta_n P_i^3},
\end{split}
\end{equation}
where $P_i$ is the launch power of channel $i$, $P_\textnormal{ASE}$ is the amplified spontaneous emission (ASE) noise power over the channel bandwidth and $\eta_n$ is the nonlinear interference coefficient after $n$ spans. The SNR is a function of the spectral location of the channel within the optical signal as $P_\textnormal{ASE}$ and $\eta_n$ are frequency dependent quantities. When the channel bandwidth $B_{\text{ch}}$ is small compared to the total optical bandwidth $B$, the power spectral density (PSD) of the NLI can be considered locally flat and $\eta_n$ can be approximated as
\begin{equation}
\begin{split}
\label{eq:eta_conversation}
\eta_n\left(f_i\right) = \int_{-\frac{B_{\text{ch}}}{2}}^{\frac{B_{\text{ch}}}{2}}\frac{G\left(\nu+f_i\right)}{P_i^{3}}d\nu \approx \frac{B_{\text{ch}}}{P_i^{3}}G\left(f_i\right),
\end{split}
\end{equation}
where $f_i$ is the center frequency of channel $i$ and $G\left(f\right)$ is the PSD of the nonlinear interference. For later use, we further define the total optical launch power as
\begin{equation}
\begin{split}
\label{eq:Ptot}
P_{\text{tot}} = \int G_{\text{Tx}}(\nu)d\nu = \sum_{\forall i} P_i,
\end{split}
\end{equation}
where $G_{\text{Tx}}$ is the input PSD of the entire optical signal. The aim of the next sections is to find an analytical expression for the nonlinear interference PSD $G\left(f\right)$, in order to compute the channel dependent SNR or similar performance metrics. 
\subsection{The nonlinear interference power}
\label{sec:NLIpower}
Various models have been proposed in the past in order to calculated the NLI power, where it is generally assumed that all frequency components attenuate in the same manner along a fiber span \cite{Splett_1993_utc,Tang_2002_tcc,Poggiolini_2012_tgm,Chen_2010_cef,Johannisson_2013_pao,Mecozzi_2012_nsl,Secondini_2012_afc,Dar_2013_pon,Carena_2014_emo,Golani_2016_mtb,Serena_2015_ate,Ghazisaeidi_2017_ato,Carena_2012_mot}. However, this assumption is no longer satisfied when transmission systems operate at large optical bandwidths (C+L band and beyond). This is because each frequency component undergoes a different power evolution during propagation as a result of ISRS and a frequency dependent attenuation coefficient. In addition, frequency dependent signal power profiles are present in hybrid amplification schemes, where part of the spectrum is amplified using distributed Raman amplifiers in order to reduce the ASE noise power for longer wavelengths or extend the amplification window beyond conventional EDFA's \cite{Cai_2015_4tt,Foursa_2002_2to,Pilipetskii_2006_hcu}. 
\par 
For a frequency dependent power evolution, the PSD of the NLI after one span is derived in Appendix \ref{derivation} and it is found to be 
\begin{equation}
\begin{split}
&G(f) = \frac{16}{27}\gamma^2\int df_1\int df_2\ G_{\text{Tx}}(f_1)G_{\text{Tx}}(f_2)G_{\text{Tx}}(f_1+f_2-f)\\
&\cdot\left|\int_0^Ld\zeta\ \sqrt{\frac{\rho(\zeta,f_1)\rho(\zeta,f_2)\rho(\zeta,f_1+f_2-f)}{\rho(\zeta,f)}}e^{j\phi\left(f_1,f_2,f,\zeta\right)}\right|^2, \\
\label{eq:general_G}
\end{split}
\end{equation}
where $\phi=-4\pi^2\beta_2\left[(f_1-f)(f_2-f)+\pi\beta_3(f_1+f_2)\right]\zeta$, $\gamma$ is the nonlinearity coefficient and $\rho(z,f)$ is the normalized signal power profile. For example, the normalized signal power profile of a passive fiber with constant attenuation coefficient $\alpha$ is $\rho(z,f)=e^{-\alpha z}$. For multi-span systems, where each span has identical fiber parameters and signal power profiles, the phased-array term 
\begin{equation}
\begin{split}
&\left|\frac{\sin\left[\frac{1}{2}n\phi\left(f_1,f_2,f,L\right)\right]}{\sin\left[\frac{1}{2}\phi\left(f_1,f_2,f,L\right)\right]}\right|^2\\
\label{eq:phasedarry}
\end{split}
\end{equation}
must be inserted into the integral in \ref{eq:general_G}. For multi-span systems with non-identical signal power profiles for each span (e.g. when non-ideal gain flattening filters are considered), the phased-array term cannot be used. Instead Eq. \eqref{eq:general_G} must be considered, where $L$ must be reinterpreted as the link length and $\rho(z,f)$ as the signal power profile for the \textit{entire} link. However, identical spans are considered for the remainder of this work.
\par 
We note that \eqref{eq:general_G} is different than the result derived in \cite[Eq.~(16)]{Roberts_17_cpo} and \cite[Eq.~(18)]{Cantono_2012_ada}, where $\rho(\zeta,f_1+f_2-f)$ and $\rho(\zeta,f)$ are swapped. \footnote{As a consequence of the different result, the $f_1$ and $f_2$ dependence vanishes, for power profiles of the form $\rho(z,f)=e^{a(z)\cdot f+b(z)}$ (as the one resulting from ISRS). This means that in the nonlinear process all three frequencies in the triplet ($f$,$f_1$,$f_2$) attenuate according to frequency $f$ which overestimates the impact of ISRS. In contrast, in Eq. \eqref{eq:general_G} each frequency ($f_1$,$f_2$,$f$) correctly attenuates with its respective power profile.} However, in section \ref{sec:validation}, it is shown by split-step simulations that \eqref{eq:general_G} is the correct formula.
\par 
Eq. \eqref{eq:general_G} can be used to model any arbitrary frequency dependent signal power profile and it is therefore suitable to evaluate the impact of ISRS on the optical Kerr nonlinearity.

\subsection{Inter-channel stimulated Raman scattering}
In the following, the PSD of the NLI in the presence of ISRS is presented, which is hereafter referred to as the ISRS GN model. In modern dispersion uncompensated links, ISRS effectively amplifies high wavelength components at the expense of low wavelength components \cite{Forghieri_1995_eom,Ho_2000_spo,Norimatsu_2001_wdd}. The resulting frequency dependent signal power profile can be obtained by solving a set of coupled ordinary differential equations \cite[Eq.~(3)]{Tariq_1993_acm}
\begin{align}
\begin{split}
&\frac{\partial P_i}{\partial z} =  \\
&- \underbrace{\displaystyle\sum_{j=i+1}^{M} \frac{1}{2}\frac{f_j}{f_i}g_{\text{r}}(\Delta f) P_j P_i}_{\textnormal{ISRS loss}} + \underbrace{\displaystyle\sum_{j=1}^{i-1} \frac{1}{2}g_{\text{r}}(\Delta f) P_j P_i}_{\textnormal{ISRS gain}} - \alpha \left(f_i\right) P_i,
\label{eq:ISRS_ODE}
\end{split}
\end{align}
where $M$ is the total number of WDM channels, $g_r(\Delta f)$ is the normalized (by the effective core area $A_{\text{eff}}$) Raman gain spectrum for a frequency separation $\Delta f=\left|f_i-f_j\right|$ and $\alpha\left(f\right)$ is the frequency dependent attenuation coefficient. The index of the channel with the highest center frequency is $i=1$. Eq. \eqref{eq:ISRS_ODE} can be extended to include distributed Raman amplification using \cite[Eq. 1]{Bromage_2004_raf}.  Eq. \eqref{eq:ISRS_ODE} has no general analytical solution and must be solved numerically. The obtained power profile can then (after normalization) be used in \eqref{eq:general_G} to yield a semi-analytical model that accurately accounts for ISRS, a frequency dependent attenuation coefficient and distributed Raman amplification. 
\par \vspace{\baselineskip}
The variation of the attenuation coefficient does typically not exceed $0.01$~dB/km across the C+L band ranging from 1530~nm to 1625~nm and might be negligible depending on accuracy and computational complexity requirements \cite{ofsfiber}. The impact of a frequency dependent attenuation on the NLI coefficient can be loosely upper bounded by assuming that \textit{every} frequency component attenuates according to the minimum in one case and according to the maximum attenuation coefficient in another case. The resulting maximum deviation in NLI coefficient is then 
\begin{equation}
\begin{split}
\Delta \eta_1 \left[\text{dB}\right] < \left(\frac{\alpha_{\text{min}}}{\alpha_{\text{max}}}\right)\left[\text{dB}\right],
\label{eq:dev_alpha}
\end{split}
\end{equation}
where $\left(\cdot\right)\left[\text{dB}\right]$ means conversion to decibel and $\alpha_{\text{min}}$ and $\alpha_{\text{max}}$ is the minimum and maximum attenuation coefficient, respectively. In \eqref{eq:dev_alpha} it was assumed that $e^{-\alpha L}\ll 1$, $\text{ln}\left(\pi^2B^2|\beta_2|/\alpha_{\text{min}}\right) \approx \text{ln}\left[\pi^2B^2|\beta_2|/\alpha_{\text{max}}\right]$ and \cite[Eq. 5]{Poggiolini_2011_amo} was used. For an attenuation deviation of $0.01$~dB/km over 95 nm, \eqref{eq:dev_alpha} yields a maximum deviation of $\Delta \eta_1 \left[\text{dB}\right] < 0.2 \ \text{dB}$. 
\par 
This contribution might be deemed negligible and the prevailing effect that causes a frequency dependent signal power profile is inter-channel stimulated Raman scattering. Eq. \eqref{eq:ISRS_ODE} can then be solved analytically when the Raman gain spectrum is assumed to be linear up to approximately $14$~THz (i.e. up to the Stokes shift). The normalized signal power profile for a spectral component $f$ is then given by \cite[Eq. (7)]{Zirngibl_1998_amo}
\begin{equation}
\begin{split}
\rho(z,f) = \frac{P_{\text{tot}}e^{-\alpha z-P_{\text{tot}}C_{\text{r}} L_{\text{eff}}f}}{\int G_{\text{Tx}}(\nu)e^{P_{\text{tot}}C_{\text{r}} L_{\text{eff}}\nu} d\nu},
\label{eq:ISRS_anal}
\end{split}
\end{equation}
where $C_{\text{r}}$ is the slope of a linear regression of the normalized Raman gain spectrum $g_r(\Delta f)$ and $L_{\text{eff}}=\frac{1-\exp\left(-\alpha z\right)}{\alpha}$. The $z$ dependence in $L_{\text{eff}}$ is suppressed for notational convenience.
\par
\begin{figure}
\includegraphics[]{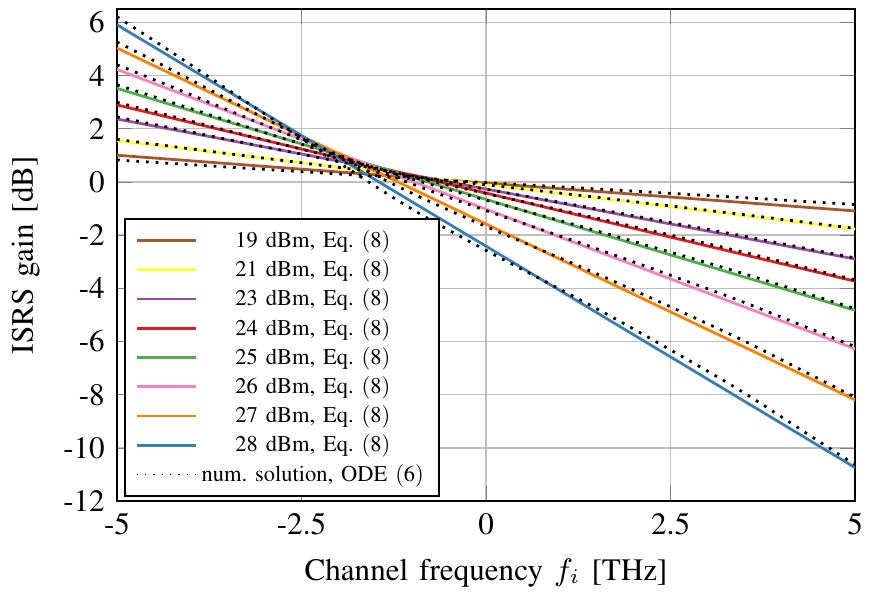}
\caption{\small The net gain due to ISRS as a function of channel frequency obtained by solving the set of coupled differential equations (6) shown in dotted lines and its analytical approximation (8) shown in solid lines for a variety of total optical launch powers $P_{\text{tot}}$.}
\label{fig:recv_power}
\end{figure}

 
The ISRS gain of a $10$~THz signal after $100$~km propagation, obtained from numerically solving \eqref{eq:ISRS_ODE} using the Raman gain spectrum as in \cite{Stolen_1973_rgi} and its analytical approximation \eqref{eq:ISRS_anal} are shown in Fig. \ref{fig:recv_power}. The precise functions that have been used can be found in \cite[Fig. 1]{Semrau_17_ard}. For a relatively high optical launch power of $28$~dBm, the average deviation between the numerical solution and its approximation is $0.18$~dB which can be considered negligible. Therefore, the analytical solution \eqref{eq:ISRS_anal} is sufficiently accurate for modeling the nonlinear interference power.
\par \vspace{\baselineskip}
Substituting \eqref{eq:ISRS_anal} in \eqref{eq:general_G} yields the reference formula of the analytical ISRS GN model \eqref{eq:ISRS_GNmodel} as 
\begin{equation}
\begin{split}
G(f) &= \frac{16}{27}\gamma^2\int df_1 \int df_2 \ G_{\text{Tx}}(f_1)G_{\text{Tx}}(f_2)G_{\text{Tx}}(f_1+f_2-f) \\
&\cdot \left| \int_0^L d\zeta \ \frac{P_{\text{tot}}e^{-\alpha \zeta-P_0C_{\text{r}} L_{\text{eff}}(f_1+f_2-f)}}{\int G_{\text{Tx}}(\nu)e^{P_{\text{tot}}C_{\text{r}} L_{\text{eff}}\nu} d\nu}e^{j\phi\left(f_1,f_2,f,\zeta\right)}\right|^2.
\label{eq:ISRS_GNmodel}
\end{split}
\end{equation}
Eq. \eqref{eq:ISRS_GNmodel} is a key result of this work which is extensively validated in Sec. \ref{sec:validation} and further applied to a C+L band case study in Sec. \ref{sec:C+Lband}. 
\par
It is useful to analyze Eq. \eqref{eq:ISRS_anal} in more detail. After trivial algebraic manipulations, we obtain that the power transfer between the spectral edges can be computed as
\begin{equation}
\begin{split}
\Delta \rho \left(z\right)\left[\text{dB}\right] = 4.3\cdot P_{\text{tot}}C_{\text{r}} L_{\text{eff}}B,
\label{eq:max_pwrspread}
\end{split}
\end{equation}
which is \textit{independent} of the spectral distribution of the input power. For modern fiber parameters $L_{\text{eff}}=26$~km, $C_{\text{r}}=0.008$~$1$/W/km/THz (approximately corresponding to a Corning$^{\tiny{\textregistered}}$ Vascade$^{\tiny{\textregistered}}$ EX2000 fiber with $A_{\text{eff}}=111$~$\mu$m) and an optimum launch power of $40$~fW/Hz (corresponding to $2$~dBm over $40$~GHz) as in \cite{Saavedra_2017_eao}, an ISRS power transfer of $\Delta \rho=1$~dB is present at a bandwidth of $5.3$~THz and $\Delta \rho >4$~dB for bandwidths larger than $10.6$~THz. 
\par 
It should be further noted that the ISRS gain is independent of the Raman gain slope and the optical launch power as long as their product is kept constant. For a given optical bandwidth, this also holds for the nonlinear interference coefficient. This will be useful for relating the results in Sections \ref{sec:validation} and \ref{sec:C+Lband} to fibers with different Raman gain slopes. 
\section{Numerical Validation}
\label{sec:validation}
In this section, the analytical ISRS GN model \eqref{eq:ISRS_GNmodel} is validated by split-step simulations for an optical fiber communication system with parameters listed in Table \ref{tab:parameters} (a). Numerically solving the Manakov equation for the entire C+L band (approximately 10 THz) is extremely challenging, due to high memory requirements and the excessive use of very large fast Fourier transforms. 
\par 
Therefore, the validation is carried out over a bandwidth of $B=1$~THz with an artificially tenfold increased Raman gain slope $C_{\text{r}}$. Using \eqref{eq:max_pwrspread}, the resulting ISRS gains for $1$~THz bandwidth coincide with the ones shown in Fig. \ref{fig:recv_power} as the product $B\cdot C_{\text{r}}$ is kept constant. The reader can therefore conveniently refer to Fig. \ref{fig:recv_power} for the ISRS gains that are present at a particular total launch power (where the abscissa must be scaled down by $10$). 
\begin{center}
\captionof{table}{System Parameters}
\label{tab:parameters}
  \begin{tabular}{ l | c | c }
    \hline
   \textbf{Parameters} & \textbf{(a)} & \textbf{(b)}  \\ \hline
   
   \textbf{for section} & \textbf{\ref{sec:validation}} & \textbf{\ref{sec:C+Lband}}  \\ \hline \hline
    Loss ($\alpha$) [dB/km]& \multicolumn{2}{c}{0.2} \\ \hline
    Dispersion ($D$) [ps/nm/km]& \multicolumn{2}{c}{17} \\ \hline
     Dispersion slope ($S$) [ps/$\text{nm}^2$/km]& 0 & 0.092 \\ \hline
    NL coefficient ($\gamma$) [1/W/km]& \multicolumn{2}{c}{1.2}\\ \hline
    Raman gain slope ($C_{\text{r}}$) [1/W/km/THz]& 0.28 & 0.028 \\ \hline
    Raman gain ($C_{\text{r}}\cdot 14$ THz) [1/W/km]& 4 & 0.4 \\ \hline
    Fiber length ($L$) [km]& \multicolumn{2}{c}{100}  \\ \hline
    Noise figure [dB]&  \multicolumn{2}{c}{5}  \\ \hline
    Symbol rate [GBd]&   10 & 50\\ \hline
      Channel spacing ($B_{\text{ch}}$) [GHz]&  10.001 &50.001 \\ \hline
     Number of channels &  101 & 201 \\ \hline
       Optical bandwidth ($B$) [THz]&  $1.01$ & $10.05$ \\ \hline
    Roll-off factor [\%]&  \multicolumn{2}{c}{0.01}  \\ \hline
    Number of symbols [$2^{x}$]& 15 & \\ \hline
    Step size [m]& 5 & \\ \hline
  \end{tabular}
\end{center}
\par 
\begin{figure*}
\centering
\includegraphics[]{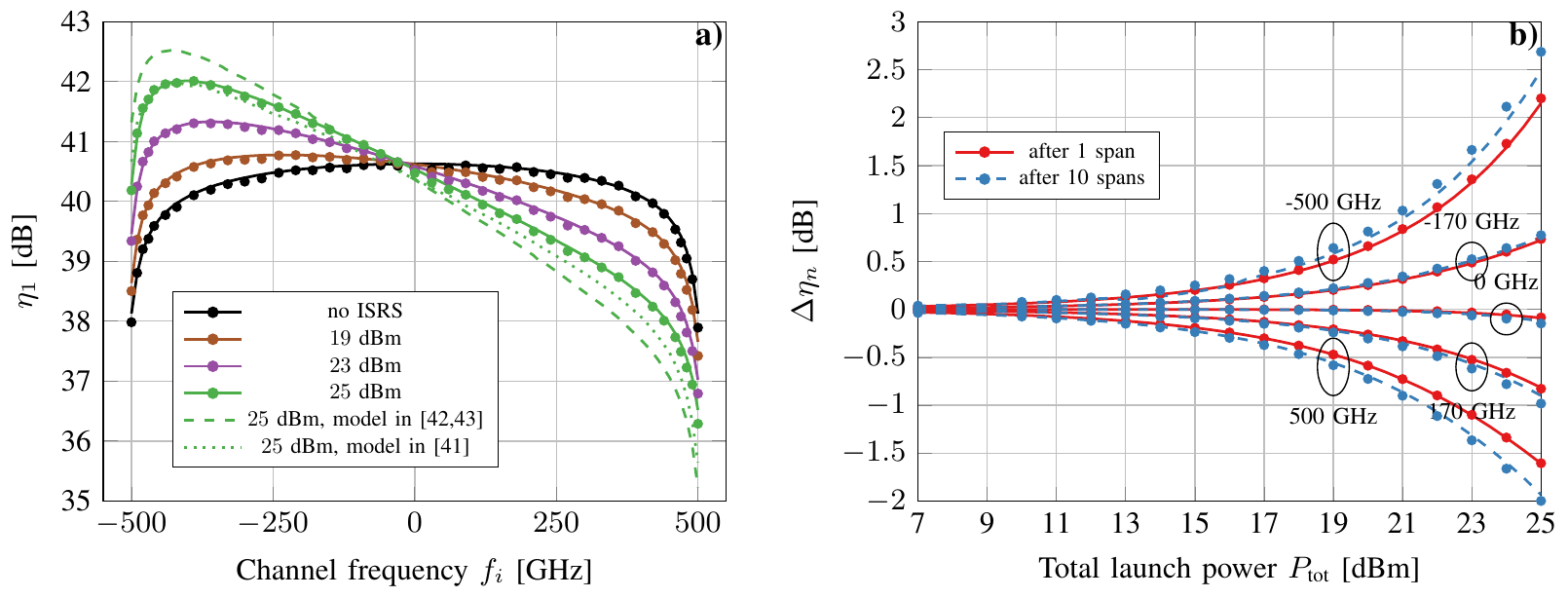}
\caption{\small The nonlinear interference coefficient after 1 span as a function of channel frequency for different total launch powers is shown in a) and the NLI change due to ISRS as a function of total launch power is shown in b). Solid lines represent the analytical ISRS GN model (9) and markers represent results obtained by split-step simulations. For comparison, the model in [41] is shown with a dotted line and the model in [42,43] is shown with a dashed line for $25$~dBm launch power in a). }
\label{fig:eta(channels)_1span_gaussian}
\end{figure*}
\begin{figure*}
\centering
\includegraphics[]{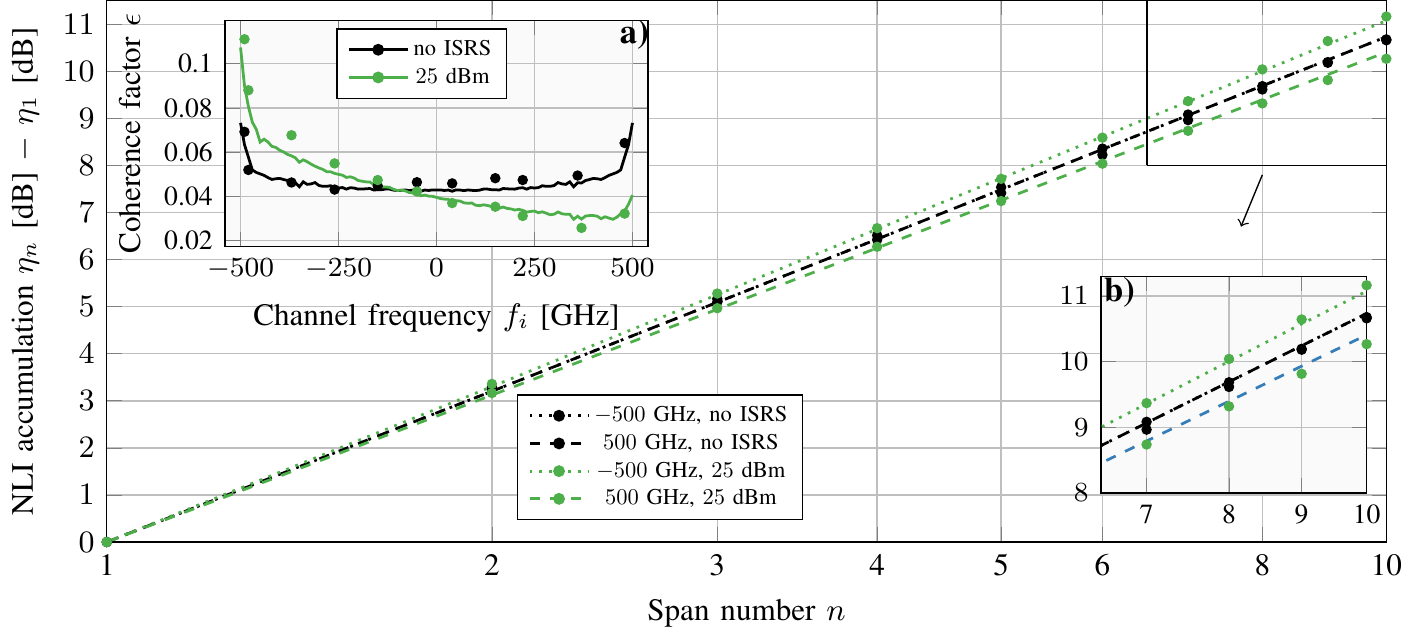}
\caption{\small The accumulation of NLI as a function of span number for the most outer frequencies of the signal with and without ISRS. Lines represent the analytical ISRS GN model (9) and markers represent split-step simulations. The inset a) shows the coherence factor after 10 spans and the inset b) shows a magnified area of the figure.}
\label{fig:eta(span)}
\end{figure*}
Gaussian modulation was implemented in order to emulate the signal Gaussianity assumption of the GN model. Additionally, a small channel bandwidth was chosen such that $B\gg B_{\text{ch}}$. A matched root-raised-cosine (RRC) filter was used to obtain the output symbols and the SNR was ideally estimated as the ratio between the variance of the transmitted symbols $E[|X|^2]$ and the variance of the noise $\sigma^2$, where $\sigma^2=E[|X-Y|^2]$ and $Y$ represents the received symbols after digital signal processing. The nonlinear interference coefficient was then estimated via \eqref{eq:SNR} and was compared with the predictions of the ISRS GN model via \eqref{eq:eta_conversation} and \eqref{eq:ISRS_GNmodel}. A spectrally uniform launch power was assumed yielding 
\begin{equation}
\begin{split}
G_{\text{Tx}}(f)=\dfrac{P_{\text{tot}}}{B}\Pi \left(\frac{f}{B}\right),
\label{eq:inputPSD}
\end{split}
\end{equation}
for the model calculations, where $\Pi\left(x\right)$ denotes the rectangular function. In the simulation environment, a frequency dependent power profile was implemented to emulate the power transfer between channels due to ISRS based on \eqref{eq:ISRS_anal}. 
\par \vspace{\baselineskip}
The results for the nonlinear interference coefficient after one span as a function of the channel frequency $f_i$ is shown in Fig. \ref{fig:eta(channels)_1span_gaussian}a) and as a function of total launch power in Fig. \ref{fig:eta(channels)_1span_gaussian}b). The accuracy of the ISRS GN model is remarkable with a maximum deviation of $<0.1$~dB. The deviation is slightly higher at the exact spectral edges. At the exact spectral edges the NLI PSD varies over the channel bandwidth and the NLI PSD cannot be considered locally flat as in \eqref{eq:eta_conversation}. This is not an inherit approximation of the ISRS GN model and it can be lifted by properly integrating over the NLI PSD. As expected, the ISRS GN model converges to the conventional GN model for low launch powers. For increasing launch powers, the nonlinear interference PSD begins to tilt. The NLI is decreased for channels that experience net ISRS loss and increased for channels that experience net ISRS gain. Moreover, as shown in Fig. \ref{fig:eta(channels)_1span_gaussian}b), the NLI interference coefficient depends exponentially on the launch power. Fig. \ref{fig:eta(channels)_1span_gaussian}b) indicates further that the launch power dependence on the NLI coefficient is stronger for an increasing number of spans.
\par 
For Gaussian modulation, the NLI accumulation in decibel as a function of fiber spans can be written as \cite[Sec. IX]{Poggiolini_2012_tgm}
\begin{equation}
\begin{split}
\left(\eta_n\right)\left[\text{dB}\right] - \left(\eta_1\right)\left[\text{dB}\right] = \left(1+\epsilon\right)\cdot \left(n\right) \left[\text{dB}\right],
\label{eq:epsilon_law}
\end{split}
\end{equation}
where $\left(\cdot \right)\left[\text{dB}\right]$ denotes conversion to decibel scale and $\epsilon$ is the coherence factor that is a measure for coherent accumulation of the NLI. As the coherence factor depends on the signal power profile (cf. \cite[Fig.~10]{Poggiolini_2012_tgm} and \cite[Fig.~3]{Semrau_2017_ace}), it is affected by ISRS. The accumulation of NLI together with the resulting coherence factor obtained from the ISRS GN model and simulation results are shown in Fig. \ref{fig:eta(span)}. Indeed, ISRS introduces a power dependent tilt on the coherent accumulation. This corresponds to the increasing power dependence of the NLI coefficient with increasing span number.  
\par 
\begin{figure*}
\centering
\includegraphics[]{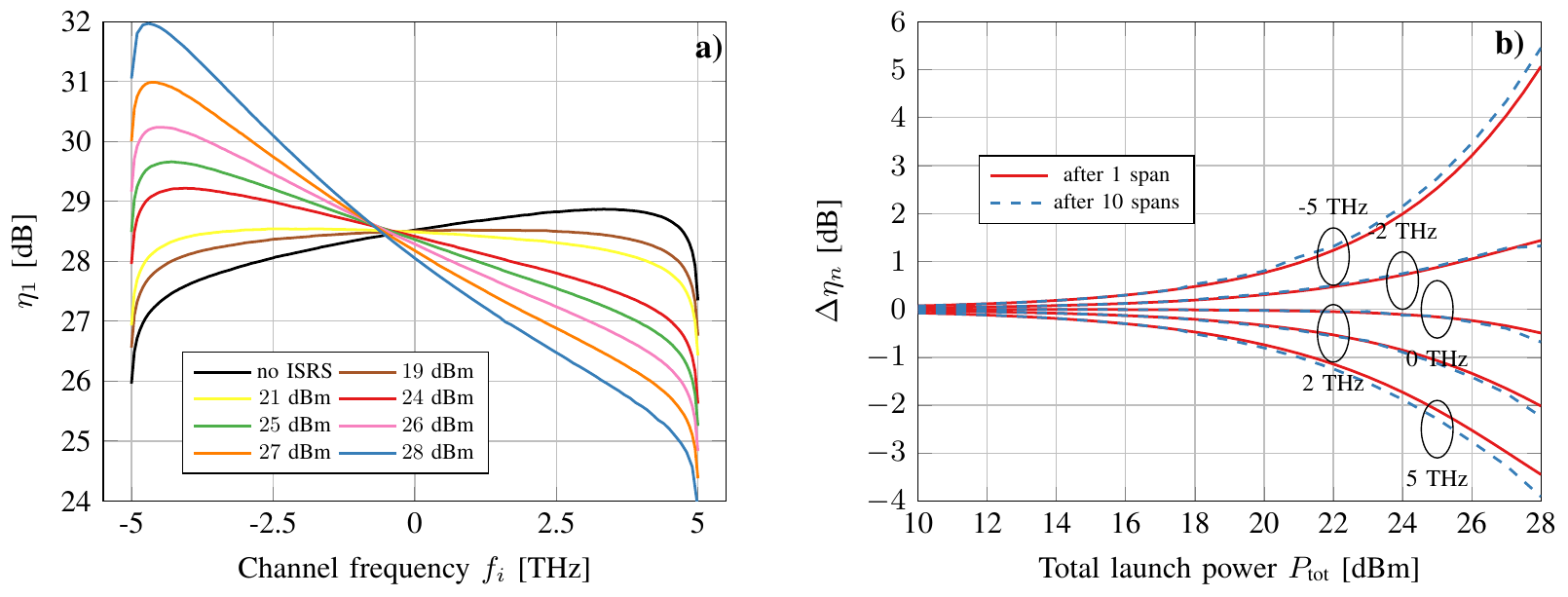}
\caption{\small The nonlinear interference coefficient after 1 span as a function of channel frequency for different total launch powers is shown in a) and the NLI as a function of total launch power is shown in b) obtained by the analytical ISRS GN model (9). The uniform optimum launch power for the system under test is $24$~dBm.}
\label{fig:NLI_1span}
\end{figure*}
\begin{figure}
\includegraphics[]{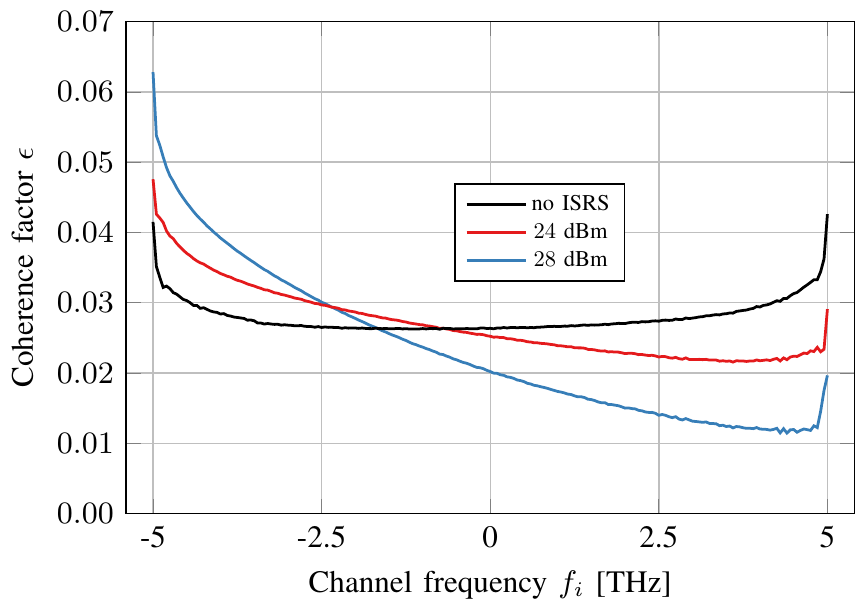}
\caption{\small The coherence factor as a function of channel frequency for a variety of total launch powers obtained by the analytical ISRS GN model (9).}
\label{fig:epsilonWB}
\end{figure}
To benchmark our results against previous works, the results in \cite[Eq.~(16)]{Roberts_17_cpo} and \cite[Eq.~(18)]{Cantono_2012_ada} are shown in dashed in Fig. \ref{fig:eta(channels)_1span_gaussian}a) using the same frequency dependent power profile \eqref{eq:ISRS_anal}. The difference originates for the reason described in Sec. \ref{sec:NLIpower}, effectively assuming that all frequencies in the nonlinear process attenuate as the one that is evaluated (i.e. $f_i$). The deviation therefore increases with increasing ISRS gain towards the spectral edges.
\par
The model in \cite{Semrau_17_ard} is shown in dotted lines. The model implements the conventional GN model, where an effective attenuation coefficient is used, that matches the effective length of the evaluated channel $f_i$. Consequently, the frequency dependent attenuation within the nonlinear process is not properly accounted for, mainly resulting in an underestimation of the ISRS impact.
\par
Bases on the numerical validation carried out in this section, it is concluded that the ISRS GN model \eqref{eq:ISRS_GNmodel} accurately predicts the nonlinear interference resulting from inter-channel stimulated Raman scattering.
\section{C+L band transmission}
\label{sec:C+Lband}

In this section, the ISRS GN model is used to evaluate the impact of ISRS on a C+L band transmission system covering $10$ THz of optical bandwidth with parameters listed in Table \ref{tab:parameters} (b). The NLI coefficient as a function of channel frequency is shown in Fig. \ref{fig:NLI_1span}a). For a particular launch power, the corresponding ISRS gains can be found in Fig. \ref{fig:recv_power}. 
\par 
Moreover, the results can be converted to a different Raman slopes coefficient by subtracting the deviation in decibel from the total launch power (cf. Eq. \eqref{eq:max_pwrspread}). For example, for halving the Raman gain slope and a total launch power of $25$~dBm, one can find the resulting NLI coefficients and ISRS gains that indicate $22$~dBm.
\par 
The tilt in NLI coefficient in the absence of ISRS is due to the dispersion slope $S$ (or $\beta_3$), where lower frequencies experience a higher amount of dispersion and therefore experience reduced NLI. As the launch power is increased, the effect of ISRS starts to balance the effect of the dispersion slope in terms of NLI. At a launch power of $22$~dBm, the NLI PSD is almost flat showing that ISRS and the dispersion slope are somewhat complementing each other in flatting the NLI spectrum. The system under test exhibits an optimum launch power of $24$~dBm, which was calculated for a flat input PSD and only considering the center channel. The resulting ISRS power transfer ranges from $-3.7$~dB to $2.9$~dB while the resulting NLI deviation due to ISRS ranges from $-1.7$~dB to $2$~dB. The slope of the NLI spectrum is $-0.24$~$\frac{\text{dB}}{\text{THz}}$ over its linear-like part from $-4$ THz to $4$ THz. 
\par 
The deviation of the NLI coefficient as a function of total launch power is shown in Fig. \ref{fig:NLI_1span}b). The NLI PSD depends exponentially on the total launch power like the ISRS gain itself and already discussed in Sec. \ref{sec:validation}. The deviation at the spectral edges of the signal is $0.5$~dB at $18$~dBm launch power. In Sec. \ref{sec:validation} it was shown that the coherence factor is changed as a result of ISRS. The same effect is seen in Fig. \ref{fig:NLI_1span}b), where the deviation of the NLI is stronger for an increased number of spans. At $24$~dBm the additional deviation of NLI is $0.1$~dB at the spectral edges after $10$ spans. 
\par 
The coherence factor as a function of channel frequency is shown in Fig. \ref{fig:epsilonWB}. The coherence factor is relatively small ($\epsilon<0.07$) due to the large bandwidth. In the absence of ISRS the average coherence factor is $0.027$. Using \eqref{eq:epsilon_law} the average coherent NLI accumulation $\epsilon \cdot \left(n\right)\left[\text{dB}\right]$ is $0.3$~dB and $0.5$~dB after 10 and 50 spans, respectively. For $24$~dBm (optimum) launch power, the maximum deviation in coherence factor is found to be $0.013$ at the spectral edges. This corresponds to a deviation in coherent NLI accumulation of $0.1$~dB and $0.2$~dB after 10 and 50 spans, respectively. The change in coherence factor due to ISRS might be deemed negligible depending on the accuracy requirements of the application. 
\par
\begin{figure}
\includegraphics[]{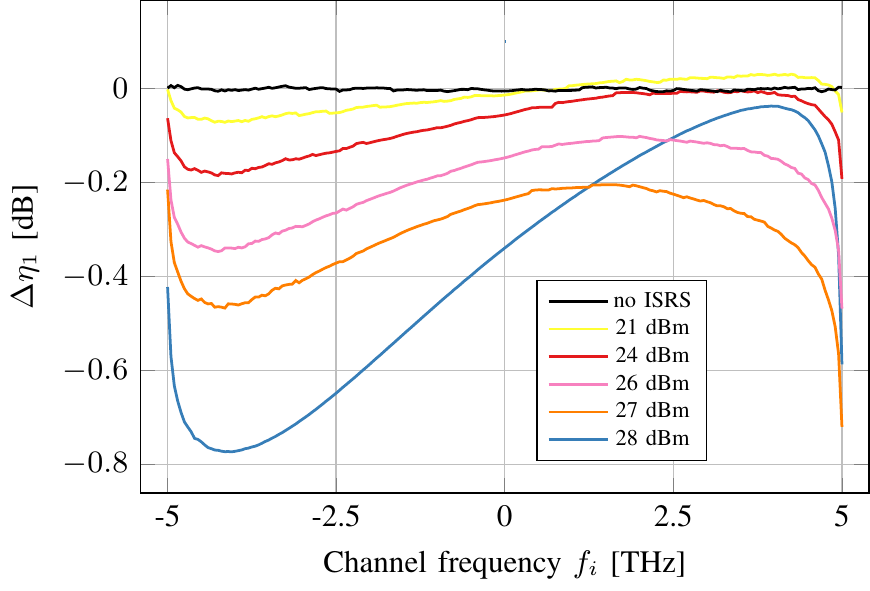}
\caption{\small Deviation of the NLI coefficient after one span between the analytical ISRS GN model (9) and \cite{Semrau_17_ard}. The validity of the ISRS GN model is shown in Sec. \ref{sec:validation}.}
\label{fig:danielWB}
\end{figure}

 
To relate our work to previously published results, the NLI coefficient after one span obtained by the ISRS GN model is compared to the works \cite{Semrau_17_ard} and \cite{Roberts_17_cpo,Cantono_2012_ada}. The signal power profile as in \eqref{eq:ISRS_anal} was used for all comparisons. The deviation of the NLI coefficient between the ISRS GN model and \cite{Semrau_17_ard} is shown in Fig. \ref{fig:danielWB}. For optimum launch power ($24$~dBm), the deviation stays below $0.19$~dB. Even for high ISRS gains at $28$~dBm launch power, the maximum deviation is $0.8$~dB.
\par
\begin{figure}
\includegraphics[]{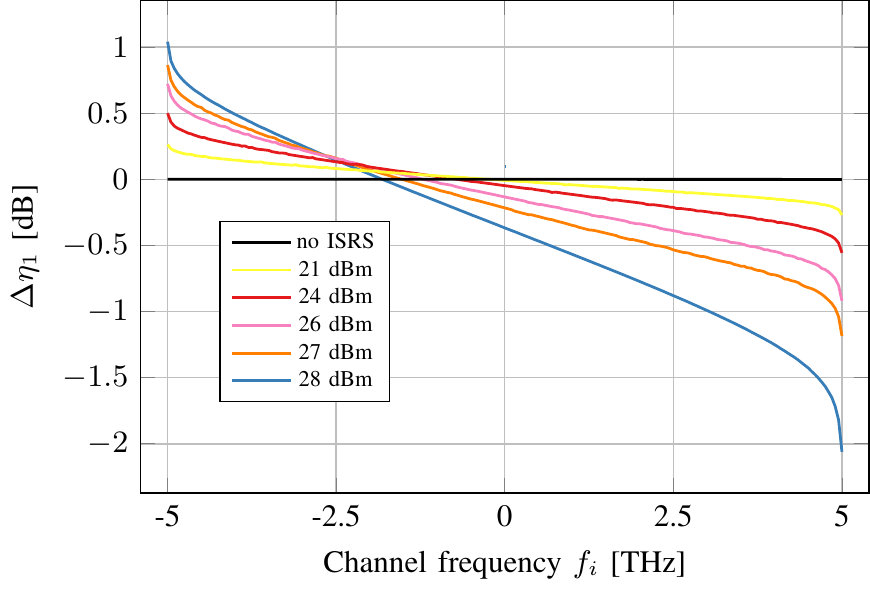}
\caption{\small Deviation of the NLI coefficient after one span between the analytical ISRS GN model (9) and \cite[Eq. (13) and (16)]{Roberts_17_cpo} \cite[Eq. (18)]{Cantono_2012_ada}. The validity of the ISRS GN model is shown in Sec. \ref{sec:validation}.}
\label{fig:standfordWB}
\end{figure}

 
The deviation of the NLI coefficient between the ISRS GN model and \cite[Eq. (13) and (16)]{Roberts_17_cpo} \cite[Eq. (18)]{Cantono_2012_ada} is shown in Fig. \ref{fig:standfordWB}. The maximum deviation is $0.56$~dB and $2.1$~dB for $24$~dBm and for $28$~dBm launch power, respectively. The reader is referred to sections \ref{sec:NLIpower} and \ref{sec:validation} for the origin of the discrepancy.
\par 
Based on the case study in this section, it is concluded that the impact of ISRS on the Kerr nonlinearity is significant in C+L band systems. This is strongly depending on launch power and the Raman gain slope due its exponential relationship to the nonlinear interference power. It should be noted that idealized gain flattening filters (GFF) were considered to compensate the ISRS power transfer at the end of each span. When realistic GFF's are considered, the ISRS gain accumulates over distance and the impact on the nonlinear interference power is more significant than as shown in this section. 
\section{Conclusion}
The ISRS GN model was introduced and presented which analytically models the impact of inter-channel stimulated Raman scattering on the nonlinear perturbation caused by Kerr nonlinearity. Its accuracy was compared to split-step simulations and a maximum deviation of $0.1$~dB in nonlinear interference power was found. The model can further account for the frequency dependent fiber attenuation, optical bandwidths beyond the Stokes shift (approximately $14$~THz) and hybrid-amplified transmission systems at the expense of greater computational complexity using a semi-analytical approach. 
\par 
It was shown that ISRS changes the nonlinear interference power by up to $2$~dB at optimum launch power for the studied C+L band transmission system. For such optical bandwidths and beyond, ISRS must be addressed in order to maximize system performance. From a physical perspective, possible solutions include the use of gain flattening filters, optimized launch power distributions or tailored fiber designs. All of which can be modeled and analyzed using the results in this paper. 
\par 
The derived ISRS GN model is therefore a powerful tool for efficient design, optimization, capacity calculations and physical-layer abstractions of ultra-wideband transmission systems that operate over the entire C+L band and beyond. 
\appendices
\section{Derivation of the ISRS GN model}
\label{derivation}
In this section, Eq. \eqref{eq:general_G} is derived for one fiber span based on the nonlinear Schr\"odinger equation (NLSE) and a first-order regular perturbation approach. The result for one span can then be extended to multiple spans using the phased-array term \eqref{eq:phasedarry} or reinterpreting a span as the entire link length with an according signal power profile. Instead of a constant attenuation coefficient $\alpha$, a generic frequency and distance dependent gain coefficient $g\left(z,f\right)$ is used to model the effect of inter-channel stimulated Raman scattering. For the sake of brevity, only the key derivation steps are outlined.
\par 
We begin with the NLSE in the frequency domain which is given by \cite[Ch.~2]{Agrawal_2012_nfo}
\begin{equation}
\begin{split}
\frac{\partial}{\partial z}E(z,f) &= \\
&\widetilde{\Gamma}(z,f) E(z,f) + j\gamma E(z,f)*E^*(z,-f)*E(z,f),
\label{eq:NLSE}
\end{split}
\end{equation}
with $\widetilde{\Gamma}(z,f)=\frac{g\left(z,f\right)}{2}+j2\pi^2\beta_2f^2+j\frac{4}{3}\pi^3\beta_3f^3$ and $u(x)*v(x)$ denoting the convolution operation. The complex envelope of the electric field $E(z,f)$ is expanded in a regular perturbation series with respect to the nonlinearity coefficient $\gamma$. The series is then truncated to first-order and we have
\begin{equation}
\begin{split}
E(z,f) =E^{\left(0\right)}(z,f)+\gamma E^{\left(1\right)}(z,f).
\label{eq:rp}
\end{split}
\end{equation}
Inserting \eqref{eq:rp} in \eqref{eq:NLSE}, we obtain
\begin{equation}
\begin{split}
E^{\left(0\right)}(z,f) =E(0,f)\cdot e^{\Gamma \left(z,f\right)},
\end{split}
\end{equation}
with $\Gamma \left(z,f\right)=\int_0^z\widetilde{\Gamma}\left(\zeta,f\right)d\zeta$ as the solution for the zeroth-order terms and a linear ordinary differential equation for the first-order terms as 
\begin{equation}
\begin{split}
\frac{\partial}{\partial z}E^{\left(1\right)}(z,f) &= \widetilde{\Gamma}(z,f) E^{\left(1\right)}(z,f) + Q(z,f),
\label{eq:rpNLSE}
\end{split}
\end{equation}
with $Q(z,f)=jE^{\left(0\right)}(z,f)*E^{\left(0\right)*}(z,-f)*E^{\left(0\right)}(z,f)$. The initial condition for the first-order solution is $E^{\left(1\right)}(0,f)=0$ and we obtain    
\begin{equation}
\begin{split}
E^{\left(1\right)}(z,f) =e^{\Gamma \left(z,f\right)}\int_0^z\frac{Q(\zeta,f)}{e^{\Gamma \left(\zeta,f\right)}}d\zeta, 
\label{eq:E1}
\end{split}
\end{equation}
as the solution of \eqref{eq:rpNLSE}.
\par 
In order to compute $Q(z,f)$, we assume that the input signal can be modeled as a periodic Gaussian process, a key assumption of the GN model, which is \cite[Eq. 13]{Carena_2012_mot}
\begin{equation}
\begin{split}
E(0,f) = \sqrt[]{f_0G_{\text{Tx}}(f)} \sum_{n=-\infty}^{\infty}\xi_n \delta\left(f-nf_0\right),
\label{eq:signal}
\end{split}
\end{equation}
where $G_{\text{Tx}}(f)$ is the power spectral density of the input signal, $\xi_n$ is a complex circular Gaussian distributed random variable, $T_0=f_0^{-1}$ is the period of the signal and $\delta(x)$ denotes the Dirac delta function. For notational convenience, we write $nf_0$ as $f_n$ and $\sum_{n=\infty}^{\infty}$ as $\sum_{\forall n}$ for the remainder of this derivation. Using \eqref{eq:signal}, $Q(z,f)$ can be written as
\begin{equation}
\begin{split}
&Q(z,f) = j f^{\frac{3}{2}}_0\sum_{\forall m}\sum_{\forall n}\sum_{\forall k}\sqrt[]{G_{\text{Tx}}(f_m)G_{\text{Tx}}(f_n)G_{\text{Tx}}(f_k)}\\
&\xi_m\xi^*_n\xi_k\delta\left(f-f_m+f_n-f_k\right)e^{\Gamma\left(z,f_m\right)+\Gamma^*\left(z,f_n\right)+\Gamma\left(z,f_k\right)}.
\label{eq:perturbed_field1}
\end{split}
\end{equation}
To first order, it can be shown that only non-degenerate frequency triplets in \eqref{eq:perturbed_field1} contribute to the nonlinear interference power. Degenerate frequency triplets merely introduce a constant phase shift of the first-order solution $E^{\left(1\right)}(z,f)$, which cancels out when the PSD of $E^{\left(1\right)}(z,f)$ is computed. For more details, the reader is referred to \cite[Ch. \rom{4}.B and \rom{4}.D]{Poggiolini_2012_ada}. Therefore, we neglect degenerate frequency triplets in order to keep the derivation concise. Similar to \cite{Poggiolini_2012_ada}, we define the triplets of non-degenerate frequency components as 
\begin{equation}
\begin{split}
A_i = \left\{(m,n,k):[m-n+k] = i\:\text{and}\:[m\neq n \:\text{or}\: k\neq n]\right\},
\end{split}
\end{equation}
and rewrite \eqref{eq:perturbed_field1} as 
\begin{equation}
\begin{split}
&Q(z,f)=j f^{\frac{3}{2}}_0 \sum_{\forall i}\delta\left(f-f_i\right)\sum_{\forall (m,n,k) \in A_i}\xi_m\xi^*_n\xi_k\\
&\sqrt[]{G_{\text{Tx}}(f_m)G_{\text{Tx}}(f_n)G_{\text{Tx}}(f_k)}e^{\Gamma\left(z,f_m\right)+\Gamma^*\left(z,f_n\right)+\Gamma\left(z,f_k\right)}.
\label{eq:perturbed_field2}
\end{split}
\end{equation}
Inserting \eqref{eq:perturbed_field2} in \eqref{eq:E1} yields the first-order solution as 
\begin{equation}
\begin{split}
&E^{\left(1\right)}(z,f) = jf^{\frac{3}{2}}_0e^{\Gamma\left(z,f\right)}\sum_{\forall i}\delta\left(f-f_i\right)\\
&\sum_{\forall (m,n,k) \in A_i}\xi_m\xi^*_n\xi_k\sqrt[]{G_{\text{Tx}}(f_m)G_{\text{Tx}}(f_n)G_{\text{Tx}}(f_k)}\\
&\int_0^z d\zeta \ e^{\Gamma\left(\zeta,f_m\right)+\Gamma^*\left(\zeta,f_n\right)+\Gamma\left(\zeta,f_k\right)-\Gamma\left(\zeta,f_m-f_n+f_k\right)} .
\label{eq:dda}
\end{split}
\end{equation}
In order to obtain the nonlinear interference power, we compute the average PSD of the first-order solution $\gamma E^{\left(1\right)}(z,f)$. Similar to \cite[Ch. \rom{4}.D]{Poggiolini_2012_ada}, the average PSD of \eqref{eq:dda} multiplied by $\gamma$ is
\begin{equation}
\begin{split}
&G_{\text{NLI}}(z,f) = 2\gamma^2 f^3_0e^{2\text{Re}\left[\Gamma\left(z,f\right)\right]}\sum_{\forall i}\delta\left(f-f_i\right)\\
&\sum_{\forall (m,n,k) \in A_i}G_{\text{Tx}}(f_m)G_{\text{Tx}}(f_n)G_{\text{Tx}}(f_k)\\
&\left|\int_0^z d\zeta \ e^{\Gamma\left(\zeta,f_m\right)+\Gamma^*\left(\zeta,f_n\right)+\Gamma\left(\zeta,f_k\right)-\Gamma\left(\zeta,f_m-f_n+f_k\right)}\right|^2 .
\label{eq:psd}
\end{split}
\end{equation}
In the following, we transform the inner summation appearing in \eqref{eq:psd} into a summation over two independent variables. For the non-degenerate set $A_i$, we have that $f_m-f_n+f_k=f_i$ and for a given frequency triplet $\left(f_i,f_m,f_k\right)$ it follows that $f_n=f_m+f_k-f_i$. Therefore, \eqref{eq:psd} can be written as
\begin{equation}
\begin{split}
&G(z,f) = 2\gamma^2 f^3_0e^{2\text{Re}\left[\Gamma\left(z,f\right)\right]}\sum_{\forall i}\delta\left(f-f_i\right)\\
&\sum_{\forall m}\sum_{\forall k}G_{\text{Tx}}(f_m)G_{\text{Tx}}(f_k)G_{\text{Tx}}(f_m+f_k-f)\\
&\left|\int_0^z d\zeta \ e^{\Gamma\left(\zeta,f_m\right)+\Gamma^*\left(\zeta,f_m+f_k-f\right)+\Gamma\left(\zeta,f_k\right)-\Gamma\left(\zeta,f\right)}\right|^2.
\label{eq:psd2}
\end{split}
\end{equation}
Finally, we define the normalized signal power profile of a frequency component as $\rho(z,f)=e^{\int_0^z g\left(\zeta,f\right) d\zeta}$ and rewrite \eqref{eq:psd2} as an integral expression by letting $f_0 \to 0$ 
\begin{equation}
\begin{split}
&G(z,f) = 2\gamma^2 \rho(z,f)\int df_1\int df_2 \  \\
&G_{\text{Tx}}(f_1)G_{\text{Tx}}(f_2)G_{\text{Tx}}(f_1+f_2-f)\\
&\left|\int_0^z d\zeta \ \sqrt[]{\frac{\rho(\zeta,f_1)\rho(\zeta,f_2)\rho(\zeta,f_1+f_2-f)}{\rho(\zeta,f)}}e^{j\phi(f_1,f_2,f,\zeta)}\right|^2 .
\label{eq:result}
\end{split}
\end{equation}
As \eqref{eq:result} was derived for single polarization, $2\gamma^2$ must be replaced by $\frac{16}{27}\gamma^2$ to obtain the nonlinear interference power for dual polarization. Furthermore, the term $\rho(z,f)$ outside of the integral can be removed when the span loss is compensated. The result is Eq. \eqref{eq:general_G}.

\section*{Acknowledgment}

The author would like to express sincere gratitude to Dr. T. Fehenberger from Technical University of Munich for fruitful discussions on the GN model derivation. Moreover, the author thanks G. Saavedra from University College London for valuable feedback on previous drafts of the paper and Dr. D.~Lavery from University College London for help with the simulation environment. Financial support from UK EPSRC programme grant UNLOC (EP/J017582/1) and a Doctoral Training Partnership (DTP) studentship to Daniel Semrau is gratefully acknowledged.

\ifCLASSOPTIONcaptionsoff
  \newpage
\fi

\bibliographystyle{IEEEtran}
\bibliography{IEEEabrv,refs/ref}

\begin{IEEEbiographynophoto}{Daniel Semrau} received the B. Sc degree in electrical engineering from the Technical University of Berlin (TUB), Berlin, Germany, in 2013, and the M. Sc. degree in Photonic Networks Engineering from Scuola Superiore Sant’Anna (SSSUP), Pisa, Italy, and Aston University, Birmingham, U.K., in 2015. In 2015 he joined the Optical Networks Group at University College London (UCL) where he is currently working towards a Ph.D. degree. His research interests are mainly focused on channel modeling, nonlinear compensation techniques and ultra-wideband transmission for long-haul coherent optical communications.
\end{IEEEbiographynophoto}

\begin{IEEEbiographynophoto}{Polina Bayvel} received the B.Sc. (Eng.) and Ph.D.
degrees in electronic and electrical engineering from University College London
(UCL), London, U.K., in 1986 and 1990, respectively. Her Ph.D. research focused
on nonlinear fiber optics and their applications.
In 1990, she was with the Fiber Optics Laboratory, General Physics Institute,
Moscow (Russian Academy of Sciences), under the Royal Society Postdoctoral
Exchange Fellowship. She was a Principal Systems Engineer with STC Submarine
Systems, Ltd., London, U.K., and Nortel Networks (Harlow, U.K., and
Ottawa, ON, Canada), where she was involved in the design and planning of
optical fiber transmission networks. During 1994–2004, she held a Royal Society
University Research Fellowship at UCL, and, in 2002, she became a Chair
in Optical Communications and Networks. She is currently the Head of the Optical
Networks Group, UCL. She has authored or co-authored more than 290
refereed journal and conference papers. Her research interests include optical
networks, high-speed optical transmission, and the study and mitigation of fiber
nonlinearities.
Prof. Bayvel is a Fellow of the Royal Academy of Engineering (F.R.Eng.),
the Optical Society of America, the U.K. Institute of Physics, and the Institute of
Engineering and Technology. She is a member of the Technical Program Committee
(TPC) of a number of conferences, including Proc. ECOC and Co-Chair
of the TPC for ECOC 2005. She was the 2002 recipient of the Institute of Physics
Paterson Prize and Medal for contributions to research on the fundamental aspects
of nonlinear optics and their applications in optical communications systems.
In 2007, she was the recipient of the Royal Society Wolfson Research
Merit Award.
\end{IEEEbiographynophoto}

\end{document}